\documentclass[aps,prb,twocolumn,superscriptaddress,showpacs]{revtex4-2}
\usepackage{color}
\usepackage{graphicx}
\usepackage{dcolumn}
\usepackage{bm}
\usepackage[mathlines]{lineno}
\usepackage[colorlinks,linkcolor=blue,anchorcolor=blue,citecolor=blue,urlcolor=blue,hyperindex,CJKbookmarks,dvipdfm]{hyperref}

\usepackage{amsmath}
\usepackage{txfonts}
\begin{document}

\title{Strain-triggered high-temperature superconducting transition in two-dimensional carbon allotrope}

\author{Tian Yan}\affiliation{Department of Physics, School of Physical Science and Technology, Ningbo University, Zhejiang 315211, China}
\author{Ru Zheng}\affiliation{Department of Physics, School of Physical Science and Technology, Ningbo University, Zhejiang 315211, China}
\author{Jin-Hua Sun}\affiliation{Department of Physics, School of Physical Science and Technology, Ningbo University, Zhejiang 315211, China}
\author{Fengjie Ma}\affiliation{The Center for Advanced Quantum Studies and Department of Physics, Beijing Normal University, Beijing 100875, China}
\author{Xun-Wang Yan}\affiliation{College of Physics and Engineering, Qufu Normal University, Shandong 273165, China}
\author{Miao Gao}\email{gaomiao@nbu.edu.cn}\affiliation{Department of Physics, School of Physical Science and Technology, Ningbo University, Zhejiang 315211, China}
\author{Tian Cui}\email{cuitian@nbu.edu.cn}\affiliation{Department of Physics, School of Physical Science and Technology, Ningbo University, Zhejiang 315211, China}
\author{Zhong-Yi Lu}\email{zlu@ruc.edu.cn}\affiliation{Department of Physics, Renmin University of China, Beijing 100872, China}

\date{\today}


\begin{abstract}
Driving non-superconducting materials into a superconducting state through specific modulation is a key focus in the field of superconductivity.
Pressure is a powerful method that can switch a three-dimensional (3D) material between non-superconducting and superconducting states.
In the two-dimensional (2D) case, strain engineering plays a similar role to pressure. However, purely strain-induced superconductivity in 2D systems remains exceedingly scarce. Using first-principles calculations, we demonstrate that a superconducting transition can be induced solely by applying biaxial tensile strain in a 2D carbon allotrope, THO-graphene, which is composed of triangles, hexagons, and octagons. Free-standing THO-graphene is non-superconducting. Surprisingly, the electron-phonon coupling in strained THO-graphene is enhanced strong enough to pair electrons and realize superconductivity, with the highest superconducting transition temperature reaching 45 K. This work not only provides a notable example of controlling metal-superconductor transition in 2D system just via strain, but also sets a new record of superconducting transition temperature for 2D elemental superconductors.
\end{abstract}

\maketitle

\section{Introduction}

Applying pressure is regarded as a clean and reversible approach to induce superconductivity in 3D quantum materials without introducing impurities.
The compound can always return to its initial non-superconducting state after the removal of pressure.
The element crystal of Li becomes superconducting under 48 GPa with superconducting transition temperature ($T_c$) reaching 20 K \cite{Shimizu-nature419}.
Black phosphorus undergoes a structural transformation from orthorhombic A17 phase to rhombohedral A7 phase at 5 GPa, accompanied by an occurrence of superconductivity \cite{Li-PNAS115}.
The stripe-type antiferromagnetic order in Mott insulator BaFe$_2$S$_3$ is destroyed by exerting pressure, with $T_c$ reaching 17 K at 13.5 GPa \cite{Takahashi-NatMater14}. The cubic alkali metal fulleride Cs$_3$C$_{60}$ can be turned into a high-$T_c$ superconductor with pressure close to 1 GPa \cite{Takabayashi-Science323}.
The fascinating discovery of novel hydrogen-rich compounds, such as H$_3$S \cite{Duan-SciRep4,Drozdov-Nature525}, LaH$_{10}$ \cite{Liu-PNAS114,Peng-PRL119,Drozdov-Nature569,Hong-CPL37}, CeH$_{10}$ \cite{Chen-PRL127}, CaH$_6$ \cite{Wang-PNAS109,Ma-PRL128,Li-NC13}, (Ca,Y)H$_6$ \cite{Liang-PRB99,Xie-JPCM31,Zhang-CPL42}, LaBe$_2$H$_8$ \cite{Zhang-PRL128,Song-PRL130}, LaB$_2$H$_8$ \cite{Song-JACS146},and LaSc$_2$H$_{24}$ \cite{He-PNAS121,Song-arXiv2510}, provides a pathway to realize room-temperature superconductivity under high-pressure condition.
Although kagome metal $A$Ti$_3$Bi$_5$ ($A$=Cs, Rb) are not superconducting at ambient pressure, double-dome superconductivity is observed in both compounds under pressure \cite{Nie-CPL42}.
Very recently, a pressure of 14 GPa triggers a structural phase transition from ambient-pressure $Amam$ phase to a high-pressure $Fmmm$ or $I4/mmm$ phase in bilayer nickelate La$_3$Ni$_2$O$_7$, resulting in a superconducting state with $T_c$ achieving 80 K at 18 GPa \cite{Sun-Nature621,Hou-CPL40,Zhang-NatPhy20}.

Acting as a role similar to pressure in the 3D case, strain engineering in 2D system can regulate the in-plane lattice parameters and related electronic properties, providing opportunity for the occurrence or enhancement of superconductivity. FeSe films deposited on SrTiO$_3$ substrate exhibit a dramatic increase of $T_c$ \cite{Wang-CPL29}, which is closely related to lattice mismatch, charge transfer, and interfacial coupling between electrons of FeSe and high-frequency optical phonons of substrate \cite{Liu-PRB85,Zhang-PRL135}.
Instead of high pressure, La$_3$Ni$_2$O$_7$, La$_{2.85}$Pr$_{0.15}$Ni$_2$O$_7$, and La$_2$PrNi$_2$O$_7$ thin films facilitated by the application of epitaxial
compressive strain show a superconducting transition at ambient pressure \cite{Ko-Nature638,Zhou-Nature640,Liu-NM24}.
Hole doping that introduced by ozone annealing and interfacial diffusion, resulting in the formation of $\gamma$ pockets at the Fermi surface, is critical to induce superconductivity \cite{Li-NSR12,Shi-CPL42}.
For infinite-layer cuprates Sr$_{1-x}$Eu$_x$CuO$_{2+y}$, the dome shape of $T_c$ versus doping level of Eu is evidently modulated by tensile strain imposed by KTaO$_3$ substrate \cite{Deng-QuantumFront4}.
Obviously, beside strain, doping effect is also of vital importance to realize or modulate superconductivity of thin film that fabricated on specific substrate. Therefore, it is quite interesting to find novel 2D materials, in which the superconducting transition can be solely controlled by strain engineering without doping.

\begin{figure}[thb]
\begin{center}
\includegraphics[width=8.6cm]{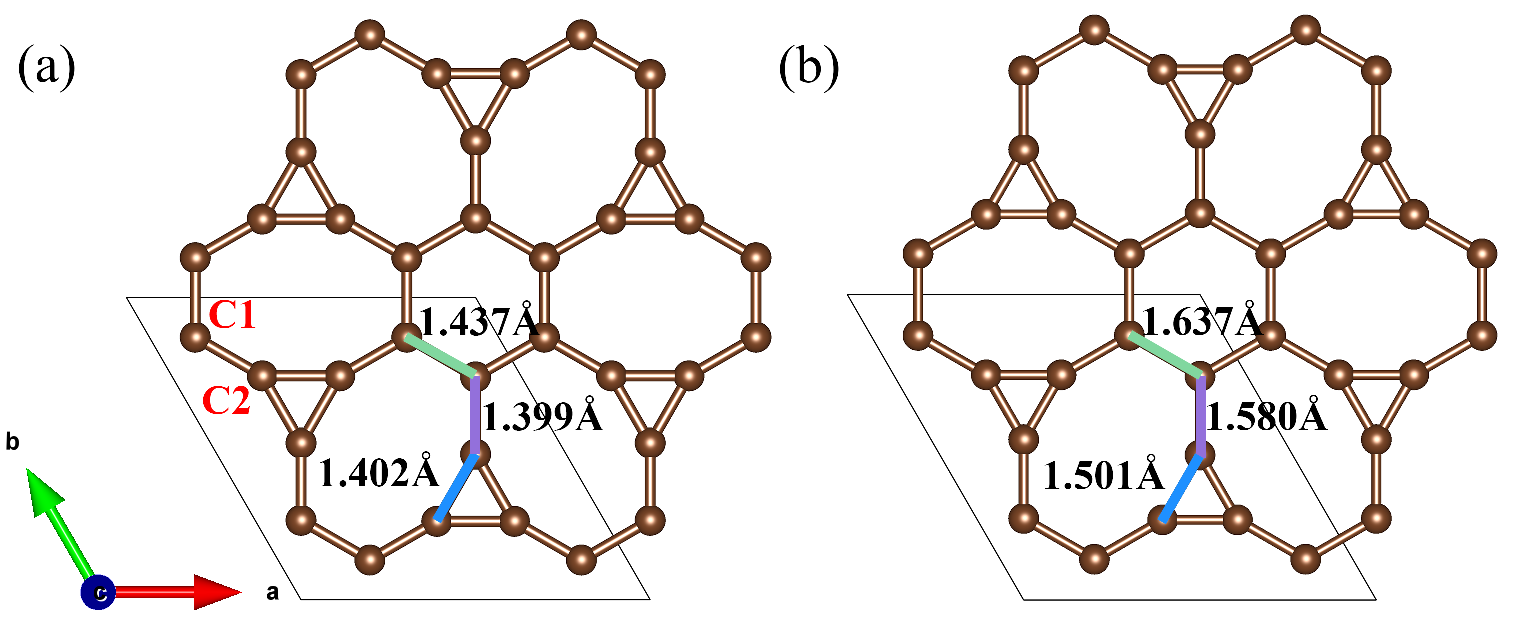}
\caption{The crystal structures of THO-graphene. (a) Free-standing THO-graphene, in which two inequivalent carbon atoms are labelled as C1 and C2.
(b) THO-graphene under biaxial tensile strain of 12\%. Three distinct C-C bonds are represented by different colors, with bond lengthes marked along the C-C bond. The unit cell is denoted by solid black lines.}
\label{fig:structure}
\end{center}
\end{figure}

In this study, we focus on a theoretically proposed 2D carbon allotrope THO-graphene, which was also named as Irida-graphene due to its similarity with the flower Iridaceae \cite{PereiraJunior-FlatChem37}. Based on the density functional theory first-principles calculations, we find that the electron-phonon coupling (EPC) of free-standing THO-graphene is too weak to induce superconductivity.
After actualizing biaxial tensile strain (BTS), EPC is gradually enhanced. THO-graphene undergoes a superconducting transition at BTS of 8\%, with $T_c$ being 1.7 K. Further increasing BTS to 12\%,  the EPC constant $\lambda$ is substantially elevated to 1.07, owing to softened phonons and enlarged density of states (DOS) at the Fermi level, accompanied by a dramatic elevation of $T_c$ to 45 K.
The detailed electronic structure, lattice dynamics, EPC, and superconducting properties for 12\%-strained THO-graphene are presented. The parameters used in our calculations are given in Appendix A.

\section{Results and discussions}

THO-graphene crystallizes in space group \textit{P}6/\textit{mmm}, where a regular hexagon is surrounded by six octagons, as shown in Fig.~\ref{fig:structure}.
For free-standing case, the in-place lattice constant is optimized as 6.315 {\AA}. The two inequivalent carbon atoms locate at Wyckoff positions 6$l$ (0.869, 0.737, 0.000) and 6$l$ (0.741, 0.481, 0.000). Three different types of carbon-carbon bonds serve to link three specific pairs of polygons: octagon with hexagon, two octagons, and octagon with triangle, respectively. The bond lengths are determined to be 1.402 {\AA}, 1.437 {\AA}, and 1.399 {\AA}, respectively. THO-graphene can withstand a maximum biaxial tensile strain of 12\% with the lattice constant being 7.073 {\AA}, before losing its dynamical stability. At this critical BTS, the coordinates of the two inequivalent carbon atoms slight shift to 6$l$ (0.866, 0.733, 0.000) and 6$l$ (0.737, 0.475, 0.000). There types of C-C bonds are elongated to 1.637 {\AA}, 1.580 {\AA}, and 1.501 {\AA}, increasing by 13.92\%, 12.94\%, and 7.06\%, respectively, with respect to the strain-free case.

\begin{figure}[htb]
\begin{center}
\includegraphics[width=8.6cm]{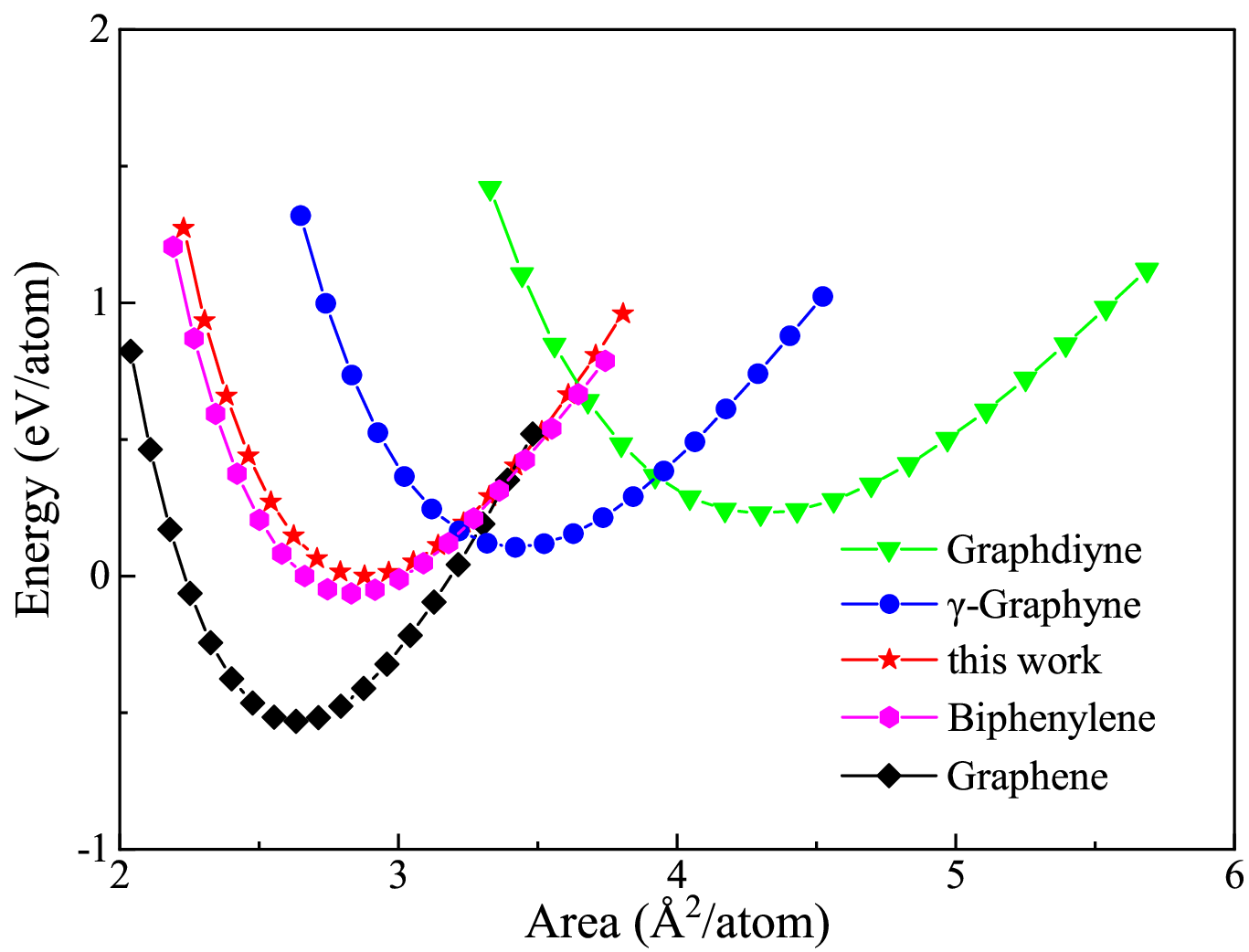}
\caption{Total energy comparison among five different 2D carbon allotropes, as a function of area per atom. Here, the total energy of optimized THO-graphene is set to zero.}
\label{fig:energy}
\end{center}
\end{figure}

\begin{figure}[t]
\begin{center}
\includegraphics[width=8.6cm]{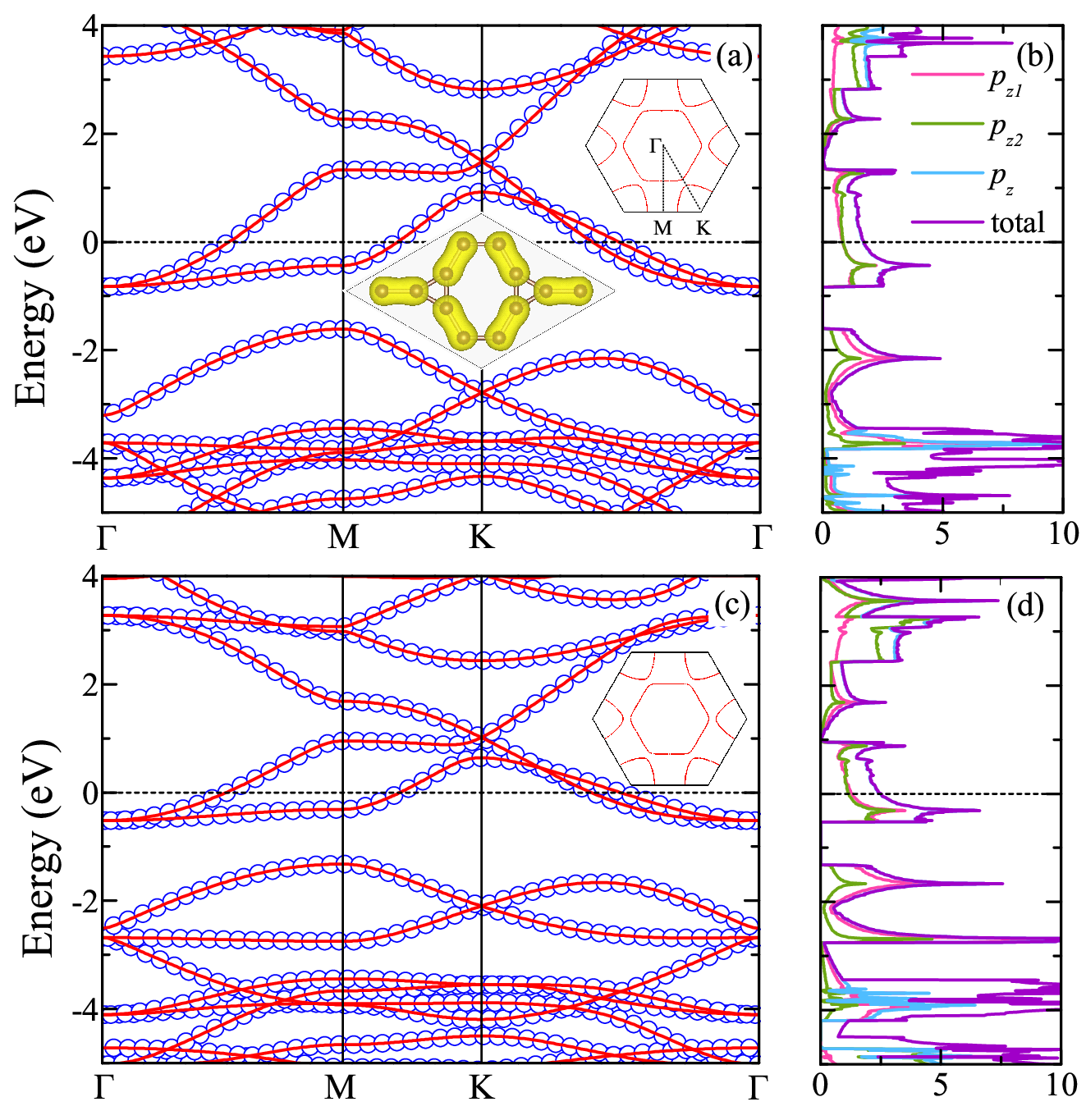}
\caption{Electronic structures of THO-graphene. (a) and (c) Band structure of THO-graphene at BTS=0 and 12\%. The red lines and blue circles denote the band structures obtained by the first-principles calculation and the MLWF interpolation, respectively. The Fermi level is set to zero. (b) and (d) Total and $p_z$-orbital-resolved DOS (states/eV/cell) of THO-graphene at BTS=0 and 12\%. $p_{z1}$ and $p_{z1}$ denote the $p_z$ orbitals of inequivalent carbon atoms marked in Fig. \ref{fig:structure}.}
\label{fig:band}
\end{center}
\end{figure}

To figure out whether it's possible to obtain THO-graphene experimentally, we compare its total energy with respect to four 2D carbon sheets that have been
synthesized [Fig. \ref{fig:energy}], i.e., graphene \cite{Novoselov-Science306}, graphdiyne \cite{Marsden-JOC70,Li-CC46}, $\gamma$-graphyne \cite{Li-Carbon136,Desyatkin-JACS144}, and biphenylene \cite{Fan-Science372}.
Although THO-graphene is less stable in comparison with graphene and biphenylene, it possesses a lower formation energy than $\gamma$-graphyne and graphdiyne, with energy advantages being 105.5 meV/atom and 228.3 meV/atom, respectively. This strongly suggests that there is a probability to acquire THO-graphene in experiment.

\begin{figure*}[t]
\begin{center}
\includegraphics[width=17.2cm]{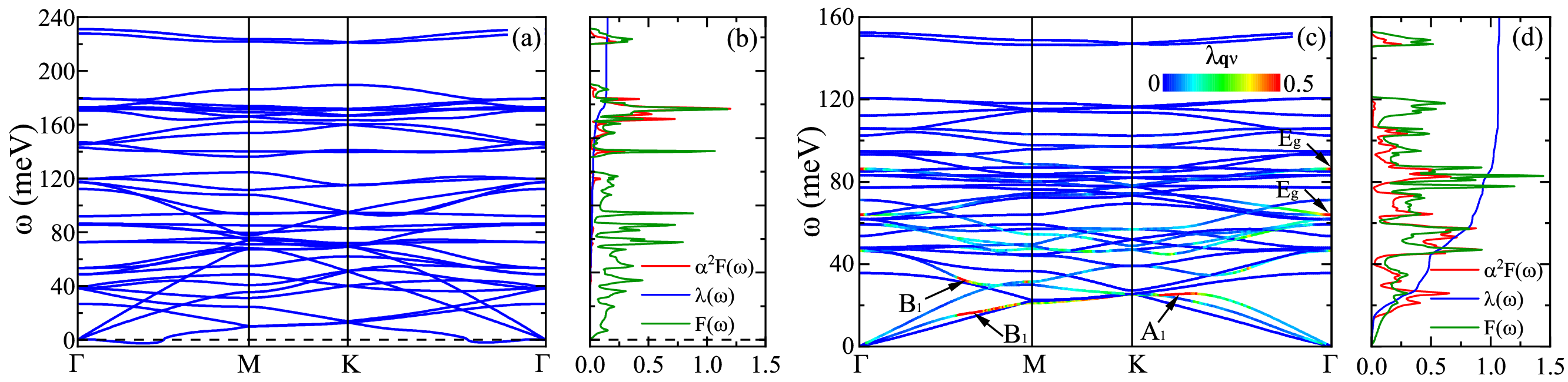}
\includegraphics[width=17.2cm]{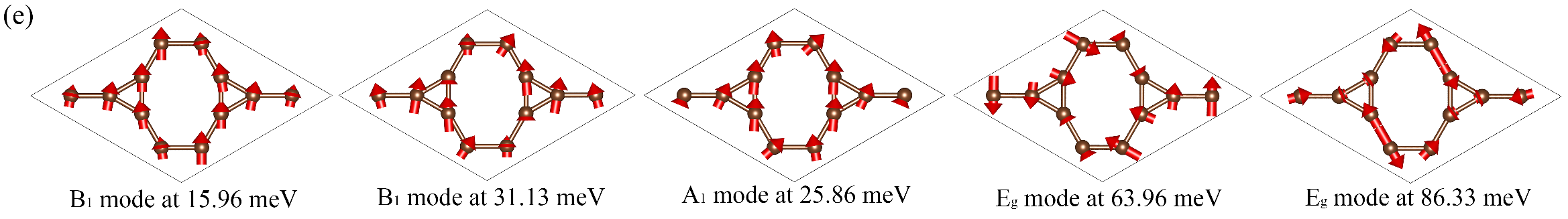}
\caption{Lattice dynamics of THO-graphene. (a) and (c) Phonon spectrum of THO-graphene at BTS=0 and 12\%. The strength of $\lambda_{\bf{q}\nu}$ for each phonon mode ${\bf{q}\nu}$ is mapped by different color. (b) and (d) Phonon density of states $F(\omega)$ (modes/meV), Eliashberg spectral function $\alpha^2 F(\omega)$, and the resulting $\lambda(\omega)$ after summation at BTS=0 and 12\%.
Here, $\lambda(\omega)=2 \int_0^{\omega'} \frac{\alpha^2 F(\omega')}{\omega'} d\omega'$.
(e) The phonon displacements of five strongly-coupled phonon modes. The length of red arrow denotes the vibrational amplitude.}
\label{fig:phonon}
\end{center}
\end{figure*}

The electronic structures of free-standing THO-graphene and those under critical BTS of 12\% are presented in Fig. \ref{fig:band}.
THO-graphene is metallic, with two energy bands across the Fermi level [Fig. \ref{fig:band}(a)], giving rise to Fermi surfaces that composed of a hexagonal electron Fermi sheet close to the $\Gamma$ point and a hole Fermi sheet around the $K$ point [see insert in Fig. \ref{fig:band}(a)]. Around the Fermi level, the DOS is dominated by $p_z$ orbital of carbon atom [Fig. \ref{fig:band}(b)]. This means that states associated with in-plane $sp^2$-hybridized $\sigma$ bonds shrink well below the Fermi level and has no contribution to conductivity. To ambiguously determine the bonding nature of THO-graphene in the vicinity of the Fermi level, we further calculate the integrated local DOS from -0.1 eV to 0.1 eV. As revealed, the Fermi surface states stem from $\pi$-bonding states surrounding the triangles.
Consequently, two inequivalent carbon atoms have almost the same weight for the DOS around the Fermi level [Fig. \ref{fig:band}(b)].
Under BTS, although the Fermi surfaces are almost unchanged [Fig. \ref{fig:band}(c)], the bandwidths are significantly narrowed, leading to an enlarged DOS [Fig. \ref{fig:band}(d)].
For example, $N(0)$, DOS at the Fermi level, increases by 45.6\% in comparison with the strain-free case, which may be beneficial to superconductivity.

The phonon spectra and DOS $F(\omega)$ for strain-free and 12\%-strained THO-graphene are given in Fig. \ref{fig:phonon}(a)-Fig. \ref{fig:phonon}(d).
Although several tiny imaginary frequencies near the $\Gamma$ point are observed in free-standing THO-graphene [Fig. \ref{fig:phonon}(a)], which disappear in the strained case [Fig. \ref{fig:phonon}(c)].
Beside those imaginary modes, other phonon modes have positive frequencies.
This phenomenon does not correspond to dynamical instability of the lattice \cite{Liu-PRB76}, and is commonly found in 2D materials simulations, such as in borophene \cite{Penev-NL16,Gao-PRB95,Gao-PRB100}, arsenene \cite{Kamal-PRB91,Kong-CPB27}, and honeycomb structures of
group-IV elements and III-V binary compounds \cite{Sahin-PRB80}.
The highest phonon frequency of unstrained THO-graphene reaches 231.05 meV, even higher than that of graphene \cite{Maultzsch-PRL92,Bonini-PRL99}.
This is probably related to the short bond length in the carbon triangle.
Due to weakening of interatomic force constants induced by BTS, the maximal phonon frequency decreases to 152.44 meV for BTS being 12\%, indicating a clear phonon softening.
Similarly, the phonon DOS peak shifts from 171.37 meV to 83.93 meV.

To investigate the EPC properties, we calculate the Eliashberg spectral function $\alpha^2 F(\omega)$.
As revealed by $\alpha^2 F(\omega)$ shown in Fig. \ref{fig:phonon}(b), only high-frequency phonons above 130 meV have contribution to the EPC in free-standing THO-graphene.
The consistency of $\alpha^2 F(\omega)$ and $F(\omega)$ in the high-frequency region indicates that the EPC matrix elements are almost uniformly distributed for these phonons.
However, high-frequency strongly-coupled phonons always lead to reduced EPC constant $\lambda$, since $\lambda=2 \int \frac{\alpha^2 F(\omega)}{\omega} d \omega$.
As a result, $\lambda$ equals to 0.15 for unstrained THO-graphene. And no superconductivity can be found by setting the Coulomb pseudopotential $\mu^*$ to 0.1.
Interestingly, $\alpha^2 F(\omega)$ is markedly red-shifted by applying 12\% BTS. Several phonon modes make significant contributions to EPC, including the acoustic $B_1$ and $A_1$ modes, and two optical $E_g$ modes around 63.96 meV and 86.33 meV [Fig. \ref{fig:phonon}(d)]. The phonon displacements of these strongly-coupled modes are depicted in Fig. \ref{fig:phonon}(e), which correspond to the in-plane vibrations of carbon atoms. The EPC constant $\lambda$ is dramatically enhanced to 1.07, about seven times that of the unstrained case, benefiting from phonon softening. Although, the logarithmic average frequency $\omega_{\text{log}}$ decreases to 39.47 meV, the superconducting $T_c$ is raised to 35.2 K, as roughly estimated by the McMillan-Allen-Dynes formula ($T_c^{\text{MAD}}$).
At intermediate BTS, the EPC constant $\lambda$, $\omega_{\text{log}}$, $T_c^{\text{MAD}}$ are calculated to be 0.35, 85.56 meV, 1.7 K for 8\% BTS, and 0.51, 66.35 meV, 10.5 K for 10\% BTS.

\begin{figure}[htbp]
\centering
\includegraphics[width=8.6cm]{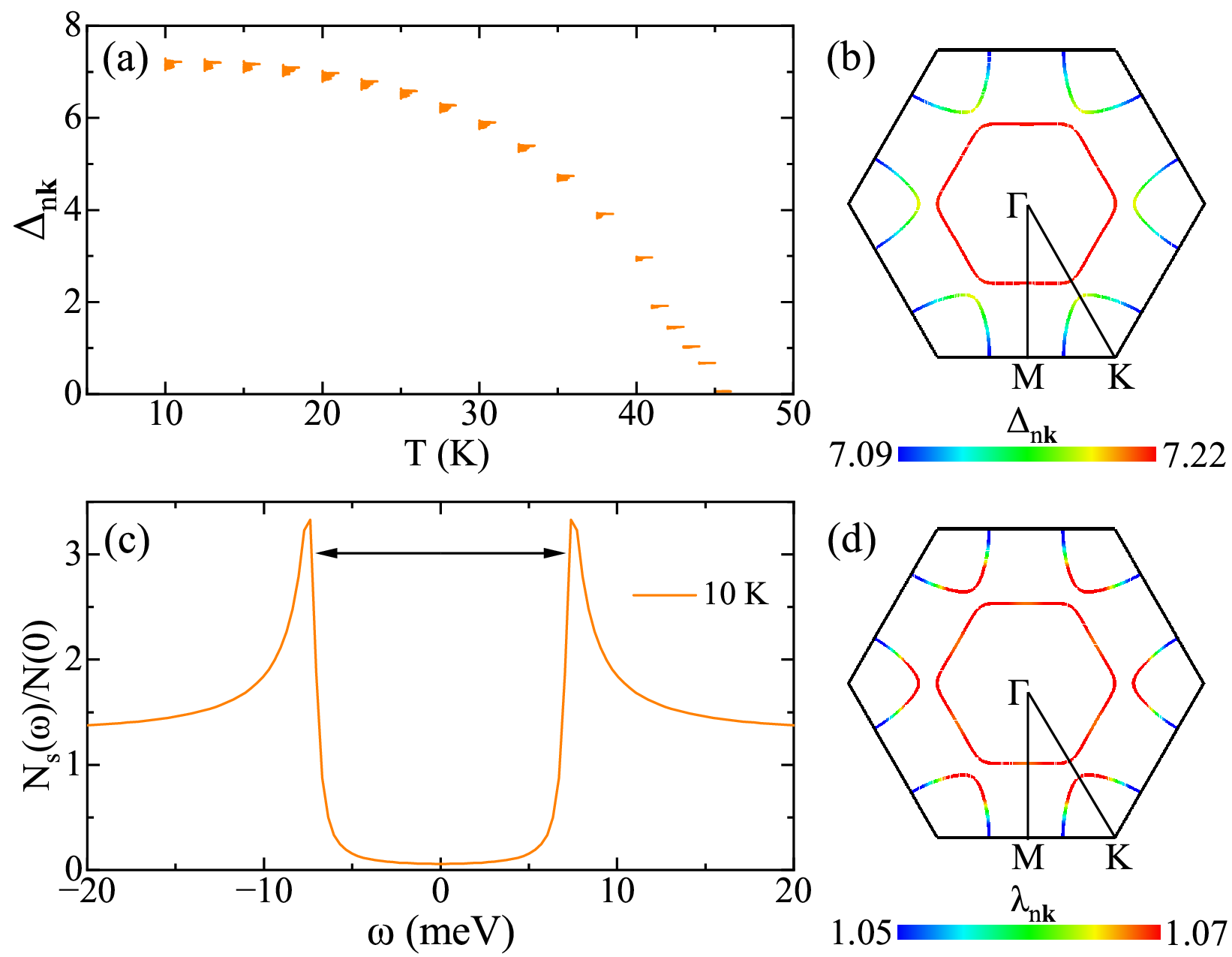}
\caption{(a) Temperature dependence of the superconducting gap values $\Delta_{n\mathbf{k}}$ on the Fermi surface at different temperatures. (b) Distribution of superconducting gap $\Delta_{n\mathbf{k}}$ on the Fermi surface at 10 K. (c) Quasiparticle density of states at 10 K. (d) Distribution of EPC strength $\lambda_{n\mathbf{k}}$ on the Fermi surface. Here, $n$ and ${\bf k}$ represent the band index and the momentum of electronic state.}
\label{fig:gap}
\end{figure}

We further solve the anisotropic Eliashberg equations to obtain a reasonable value of $T_c$ and have a insight into the distribution of superconducting gaps for BTS being 12\%. Here, the truncated frequency $\omega_c$ for the sum over Matsubara frequencies is selected as 1.5 eV, about 10 times that of the highest phonon frequency of THO-graphene under BTS of 12\%.
Figure \ref{fig:gap}(a) shows the normalized superconducting gap $\Delta_{n\mathbf{k}}$ distribution at different temperatures.
The superconducting energy gaps grouped together, suggesting a single-gap nature.
The highest temperature with nonvanished $\Delta_{n\mathbf{k}}$, i.e. $T_c$, is calculated to be 45 K.
At 10 K, the average superconducting gap $\Delta_{n\mathbf{k}}^{\text{ave}}$ for THO-graphene is 7.17 meV.
The anisotropy ratio of superconducting gap,
$\Delta^{\text{aniso}}_{n\mathbf{k}}$, defined by ($\Delta^{\text{max}}_{n\mathbf{k}}$-$\Delta^{\text{min}}_{n\mathbf{k}}$)/$\Delta^{\text{ave}}_{n\mathbf{k}}$=(7.22-7.09)/7.17=1.81\%.
The single-gap characteristic is also confirmed by the quasiparticle density of states $N_s(\omega)/N(0)$ [see Fig. \ref{fig:gap}(c)].
The distribution of superconducting gap $\Delta_{n\mathbf{k}}$ on the Fermi surface at 10 K is given in Fig. \ref{fig:gap}(b), in which the gaps on hexagon-shaped Fermi surface around the $\Gamma$ point are slightly larger. This is also the case for the distribution of $\lambda_{n\mathbf{k}}$ on the Fermi surface [Fig. \ref{fig:gap}(d)].

For theoretical predictions, strain-induced superconducting transition from a non-superconducting phase is rarely reported.
Strain modulation plays a role to enhance the $T_c$ of 2D intrinsic superconductors that proposed theoretically.
The transition temperature of monolayer MgB$_2$ is boosted from 20 K to 50 K under 4\% tensile strain \cite{Bekaert-PRB96}.
After hydrogenation, the $T_c$ of monolayer MgB$_2$ is further raised to 67 K due to the reduction of electron occupation in boron $\sigma$-bands, and up to 100 K with strain tuning \cite{Bekaert-PRL123}.
Sandwich-like trilayer films, formed by two hexagonal BC sheets and a intercalation metal layer in between, e.g. LiB$_2$C$_2$ and NaB$_2$C$_2$, are promising 2D high-$T_c$ superconductors with the highest $T_c$ achieving 150 K under BTS \cite{Gao-PRB101,Zhang-PRB110,Chen-CPL41,Liu-NL23}.
Strain heightens the EPC of hole-doped hexagonal monolayer BN, leading a maximal $T_c$ of 41.6 K at 17.5\% strain \cite{Jin-PRB101}.
Three-gap superconductivity was suggested in hydrogenated LiBC with $T_c$ above 120 K under 3.59\% tensile strain \cite{Liu-PRR6}.
Hydrogenated Janus 2H-MoSH displays an intrinsic $T_c$ of 26.8 K, tunable to 36.7 K via strain and doping \cite{Ku-PRB107}.
Thus, THO-graphene can be regarded as an outstanding platform in which metal-superconductor transition is effectively controlled by applying strain.

Finding high-$T_c$ 2D elemental superconductor is an interesting topic. Currently, superconductivity in 2D carbon sheets was extensively explored. It was reported that twisted bilayer graphene resides in an insulating state, but develops superconductivity with $T_c$ to 1.7 K under electrostatic doping \cite{Cao-Nature556}. According to a theoretical calculation, the $T_c$ of graphene reaches 30 K under hole doping of $\sim$4$\times$10$^{14}$ cm$^{-2}$ and 16.5\% BTS \cite{Si-PRL111}. T-graphene, a carbon sheet with 4- and 8-membered rings, also named as Octagraphene, was suggested to be an intrinsic elemental superconductor with $T_c$ up to 20.8 K \cite{Gu-CPL36}. By disrupting the perfect Fermi surface nesting and long-range magnetic order via electron doping, Octagraphene may exhibit unconventional $s^{\pm}$ superconductivity based on
spin fluctuation \cite{Kang-PRB99,Yao-CPL42}. Conventional and unconventional superconductivity in biphenylene was investigated, with $T_c$ being 3.0 K and 1.7 K, respectively \cite{Ye-CPL40}.
Other theoretically studied 2D element superconductor, including silicene \cite{Wan-EurophysLett104}, phosphorene \cite{Ge-NewJPhys17}, borophene \cite{Penev-NL16,Gao-PRB95,Gao-PRB100}, and arsenene \cite{Kong-CPB27}, were predicted with $T_c$ no more than 31 K. In experiment, growing thin film structure through 3D superconducting noble metal is another strategy to obtain 2D elemental superconductor, for example, Nb \cite{Zhang_CPB32}, Pb \cite{Guo-Science306,Nishio-PRL101}, and Ga \cite{Zhang-PRL114} thin films. But the transition temperatures somehow resemble those of the bulk materials. It is clear that the $T_c$ of THO-graphene predicted in this study sets a new record for 2D elemental superconductor.

\section{Conclusion}

In summary, this study demonstrates the remarkable potential of strain engineering as a powerful and clean tool to induce superconductivity in two-dimensional elemental materials. Through systematic first-principles calculations, we have shown that THO-graphene, a metallic carbon monolayer composed of triangles, hexagons, and octagons, undergoes a dramatic transition from a non-superconducting state to a robust superconductor under applied biaxial tensile strain. The emergence of superconductivity is primarily driven by the synergistic effects of strain-induced electronic and phononic modifications: a significant increase in the DOS at the Fermi level and a strong softening of specific in-plane vibrational modes. These changes collectively enhance the electron-phonon coupling strength, enabling Cooper pair formation.
Notably, at a strain of 12\%, the calculated $T_c$ reaches 45 K, which represents the highest predicted temperature among purely elemental 2D superconductors reported to date. This finding sets a new benchmark and highlights the unique responsiveness of carbon-based nanostructures to mechanical deformation. Furthermore, the energetic feasibility of THO-graphene relative to other synthesized carbon allotropes suggests promising prospects for its experimental realization.

\begin{acknowledgments}
This work was supported by the National Key R\&D Pro-gram of China (Grants No. 2024YFA1408601 and No. 2023YFA1406201), the National Natural Science Foundation of China (Grants No. 12434009, No. 12274255, No. 12074040, and No. 12104247), Zhejiang Provincial Natural Science Foundation of China (Grants No. LMS26A040005 and No. LMS26A040006), Program for Science and Technology lnnovation Team in Zhejiang (Grant No. 2021R01004), and Ningbo Young Scientific and Technological Innovation Leaders Program (Grant No. 2023QL016). F.M was also supported by the BNU Tang Scholar.
\end{acknowledgments}

\appendix

\section{Calculation Methods}

In our calculations, the first-principles plane-wave package, QUANTUM-ESPRESSO, is adopted \cite{Giannozzi-JPhysCondensMatter21}.  A vacuum layer of 20 {\AA} is added along the $c$ axis to avoid the coupling between adjacent THO-graphene sheets. We calculate the electronic states by using the generalized gradient approximation (GGA) of Perdew-Burke-Ernzerhof formula \cite{Perdew-PRL77} and the optimized norm-conserving Vanderbilt pseudopotentials \cite{Hamann-PRB88}. After full convergence test, the kinetic energy cutoff and the charge density cutoff are set to 80 Ry and 320 Ry, respectively. The charge densities are determined self-consistently on an unshifted mesh of 24$\times$24$\times$1 points with a Methfessel-Paxton smearing \cite{Methfessel-PRB40} of 0.02 Ry. The dynamical matrices and perturbation potentials are calculated on a $\Gamma$-centered mesh of 8$\times$8$\times$1 points, within the framework of density-functional perturbation theory \cite{Baroni-RMP73}.

The maximally localized Wannier functions (MLWFs) \cite{Pizzi-JPCM32} of strained THO-graphene are constructed on a 8$\times$8$\times$1 grid within the Brillouin zone. We used thirty Wannier functions to describe the band structure of THO-graphene around the Fermi level. More specifically, 12 Wannier functions correspond to the $p_z$ orbitals of carbon atoms, other 18 Wannier functions are $\sigma$-like states localized in the middle of carbon-carbon bonds. After minimization, the maximal spatial extension of these MLWFs is smaller than 1.1 {\AA}$^2$, showing excellent localization. Fine electron (160$\times$160$\times$1) and phonon (80$\times$80$\times$1) grids were employed to interpolate the EPC constant with the Wannier90 and EPW codes \cite{Lee-npjCM9}. Dirac $\delta$-functions for electrons and phonons were replaced by smearing functions with widths of 20 and 0.5 meV, respectively. The superconducting properties of THO-graphene at BTS=12\% are examined by solving the anisotropic Eliashberg equations \cite{Lee-npjCM9} using 80$\times$80$\times$1 mesh for both electrons and phonons.


\begin{references}
\bibitem{Shimizu-nature419} K. Shimizu, H. Ishikawa, D. Takao, T. Yagi, and K. Amaya,
Superconductivity in compressed lithium at 20 K, \href{https://doi.org/10.1038/nature01098}{Nature (London) \textbf{419}, 597 (2002)}.


\bibitem{Li-PNAS115} X. A. Li, J. Sun, P. Shahi, M. Gao, A. H. MacDonald, Y. Uwatoko, T. Xiang, J. B. Goodenough, J. Cheng, and J. Zhou,
Pressure-induced phase transitions and superconductivity in a black phosphorus single crystal,
\href{https://doi.org/10.1073/pnas.1810726115}{Proc. Natl. Acad. Sci. U.S.A. \textbf{115}, 9935 (2018)}.

\bibitem{Takahashi-NatMater14} H. Takahashi, A. Sugimoto, Y. Nambu, T. Yamauchi, Y. Hirata, T. Kawakami, M. Avdeev, K. Matsubayashi, F. Du, C. Kawashima, H. Soeda, S. Nakano, Y. Uwatoko, Y. Ueda, T. J. Sato, and K. Ohgushi, Pressure-induced superconductivity in the iron-based ladder material BaFe$_2$S$_3$, \href{https://doi.org/10.1038/nmat4351}{Nat. Mater. \textbf{14}, 1008 (2015)}.

\bibitem{Takabayashi-Science323} Y. Takabayashi, A. Y. Ganin, P. Jegli\v{c}, D. Ar\v{c}on, T. Takano, Y. Iwasa, Y. Ohishi, M. Takata, N. Takeshita, K. Prassides, and M. J. Rosseinsky, The Disorder-Free Non-BCS Superconductor Cs$_3$C$_{60}$ Emerges from an Antiferromagnetic Insulator Parent State, \href{https://doi.org/10.1126/science.1169163}{Science \textbf{323}, 1585 (2009)}.

\bibitem{Duan-SciRep4} D. F. Duan, Y. X. Liu, F. B. Tian, D. Li, X. L. Huang, Z. L. Zhao, H. Y. Yu, B. Liu, W. J. Tian, and T. Cui, Pressure-induced metallization of dense (H$_2$S)$_2$H$_2$ with high-$T_c$ superconductivity, \href{https://doi.org/10.1038/srep06968}{Sci. Rep. \textbf{4}, 6968 (2014)}.


\bibitem{Drozdov-Nature525} A. P. Drozdov, M. I. Eremets, I. A. Troyan, V. Ksenofontov, and S. I. Shylin, Conventional superconductivity at 203 kelvin at high pressures in the sulfur hydride system, \href{https://doi.org/10.1038/nature14964}{Nature (London) \textbf{525}, 73 (2015)}.

\bibitem{Liu-PNAS114} H. Liu, I. I. Naumov, R. Hoffmann, N. W. Ashcroft, and R. J. Hemley, Potential high-$T_c$ superconducting lanthanum and yttrium hydrides at high pressure, \href{https://doi.org/10.1073/pnas.1704505114}{Proc. Natl. Acad. Sci. U.S.A. \textbf{114}, 6990 (2017)}.


\bibitem{Peng-PRL119} F. Peng, Y. Sun, C. J. Pickard, R. J. Needs, Q. Wu, and Y. M. Ma, Hydrogen Clathrate Structures in Rare Earth Hydrides at High Pressures: Possible Route to Room-Temperature Superconductivity, \href{https://doi.org/10.1103/PhysRevLett.119.107001}{Phys. Rev. Lett. \textbf{119}, 107001 (2017)}.

\bibitem{Drozdov-Nature569} A. P. Drozdov, P. P. Kong, V. S. Minkov, S. P. Besedin, M. A. Kuzovnikov, S. Mozaffari, L. Balicas, F. F. Balakirev, D. E. Graf, V. B. Prakapenka, E. Greenberg, D. A. Knyazev, M. Tkacz, and M. I. Eremets, Superconductivity at 250 K in lanthanum hydride under high pressures, \href{https://doi.org/10.1038/s41586-019-1201-8}{Nature (London) \textbf{569}, 528 (2019)}.

\bibitem{Hong-CPL37} F. Hong, L. X. Yang, P. F. Shan, P. T. Yang, Z. Y. Liu, J. P. Sun, Y. Y. Yin, X. H. Yu, J. G. Cheng, and Z. X. Zhao,
Superconductivity of Lanthanum Superhydride Investigated Using the Standard Four-Probe Configuration under High Pressures,
\href{https://doi.org/10.1088/0256-307X/37/10/107401}{Chin. Phys. Lett. \textbf{37}, 107401 (2020)}.

\bibitem{Chen-PRL127} W. H. Chen, D. V. Semenok, X. L. Huang, H. Y. Shu, X. Li, D. F. Duan, T. Cui, and A. R. Oganov,
High-Temperature Superconducting Phases in Cerium Superhydride with a $T_c$ up to 115 K below a Pressure of 1 Megabar,
\href{https://doi.org/10.1103/PhysRevLett.127.117001}{Phys. Rev. Lett. \textbf{127}, 117001 (2021)}.

\bibitem{Wang-PNAS109} H. Wang, J. S. Tse, K. Tanaka, T. Iitaka, and Y. Ma,
Superconductive sodalite-like clathrate calcium hydride at high pressures,
\href{https://doi.org/10.1073/pnas.1118168109}{Proc. Natl. Acad. Sci. U.S.A. \textbf{109}, 6463 (2012)}.

\bibitem{Ma-PRL128} L. Ma, K. Wang, Y. Xie, X. Yang, Y. Y. Wang, M. Zhou, H. Liu, X. H. Yu, Y. S. Zhao, H. B. Wang, G. T. Liu, and Y. Ma,
High-Temperature Superconducting Phase in Clathrate Calcium Hydride CaH$_6$ up to 215 K at a Pressure of 172 GPa,
\href{https://doi.org/10.1103/PhysRevLett.128.167001}{Phys. Rev. Lett. \textbf{128}, 167001 (2022)}.


\bibitem{Li-NC13} Z. W. Li, X. He, C. L. Zhang, X. C. Wang, S. J. Zhang, Y. T. Jia, S. M. Feng, K. Lu, J. F. Zhao, J. Zhang, B. S. Min, Y. W. Long, R. C. Yu, L. H. Wang, M. Y. Ye, Z. S. Zhang, V. Prakapenka, S. Chariton, P. A. Ginsberg, J. Bass, S. H. Yuan, H. Z. Liu, and C. Q. Jin,
Superconductivity above 200 K discovered in superhydrides of calcium,
\href{https://doi.org/10.1038/s41467-022-30454-w}{Nat. Commun. \textbf{13}, 2863 (2022)}.


\bibitem{Liang-PRB99} X. W. Liang, A. Bergara, L. Y. Wang, B. Wen, Z. S. Zhao, X. F. Zhou, J. L. He, G. Y. Gao, and Y. J. Tian,
Potential high-$T_c$ superconductivity in CaYH$_{12}$ under pressure,
\href{https://doi.org/10.1103/PhysRevB.99.100505}{Phys. Rev. B \textbf{99}, 100505 (2019)}.

\bibitem{Xie-JPCM31} H. Xie, D. F. Duan, Z. J. Shao, H. Song, Y. C. Wang, X. H. Xiao, D. Li, F. B. Tian, B. B. Liu, and T. Cui,
High-temperature superconductivity in ternary clathrate YCaH$_{12}$ under high pressures,
\href{https://doi.org/10.1088/1361-648X/ab09b4}{J. Phys.: Condens. Matter \textbf{31}, 245404 (2019)}.

\bibitem{Zhang-CPL42} K. X. Zhang, J. N. Guo, Y. L. Wang, X. Y. Wu, X. L. Huang, and T. Cui,
Robust Superconducting Stability of Ternary Hydride $Im\bar{3}m$ (Y, Ca)H$_6$ upon Decompression,
\href{https://doi.org/10.1088/0256-307X/42/11/110704}{Chin. Phys. Lett. \textbf{42}, 110704 (2025)}.

\bibitem{Zhang-PRL128} Z. H. Zhang, T. Cui, M. J. Hutcheon, A. M. Shipley, H. Song, M. Y. Du, V. Z. Kresin, D. F. Duan, C. J. Pickard, and Y. S. Yao,
Design Principles for High-Temperature Superconductors with a Hydrogen-Based Alloy Backbone at Moderate Pressure,
\href{https://doi.org/10.1103/PhysRevLett.128.047001}{Phys. Rev. Lett. \textbf{128}, 047001 (2022)}.

\bibitem{Song-PRL130} Y. G. Song, J. K. Bi, Y. Nakamoto, K. Shimizu, H. Y. Liu, B. Zou, G. T. Liu, H. B. Wang, and Y. M. Ma,
Stoichiometric Ternary Superhydride LaBeH$_8$ as a New Template for High-Temperature Superconductivity at 110 K under 80 GPa,
\href{https://doi.org/10.1103/PhysRevLett.130.266001}{Phys. Rev. Lett. \textbf{130}, 266001 (2023)}.


\bibitem{Song-JACS146} X. X. Song, X. K. Hao, X. D. Wei, X. L. He, H. Y. Liu, L. Ma, G. T. Liu, H. B. Wang, J. Y. Niu, S. J. Wang, Y. P. Qi, Z. Y. Liu, W. T. Hu, B. Xu, L. Wang, G. Y. Gao, and Y. J. Tian,
Superconductivity above 105 K in Nonclathrate Ternary Lanthanum Borohydride below Megabar Pressure,
\href{https://doi.org/10.1021/jacs.3c14205}{J. Am. Chem. Soc. \textbf{146}, 13797 (2024)}.


\bibitem{Song-arXiv2510} Y. G. Song, C. H. Ma, H. B. Wang, M. Zhou, Y. P. Qi, W. Z. Cao, S. R. Li, H. Y. Liu, G. T. Liu, and Y. M. Ma,
Room-Temperature Superconductivity at 298 K in Ternary La-Sc-H System at High-pressure Conditions,
\href{https://doi.org/10.48550/arXiv.2510.01273}{arXiv:2510.01273 (2024)}.


\bibitem{He-PNAS121} X. L. He, W. B. Zhao, Y. Xie, A. Hermann, R. J. Hemley, H. Y. Liu, and Y. M. Ma,
Predicted hot superconductivity in LaSc$_2$H$_{24}$ under pressure,
\href{https://doi.org/10.1073/pnas.2401840121}{Proc. Natl. Acad. Sci. U.S.A. \textbf{121}, e2401840121 (2024)}.

\bibitem{Nie-CPL42} J. Y. Nie, X. F. Yang, K. Y. Chen, X. Q. Liu, W. Xia, J. Wang, R. Zhang, D. Z. Dai, C. C. Zhao, C. P. Tu, H. L. Dong, X. B. Jin, J. K. Deng, X. Zhang, Y. F. Guo, and S. Y. Li, Pressure-Induced Double-Dome Superconductivity in Kagome Metals ATi$_3$Bi$_5$ (A = Cs, Rb),
\href{https://doi.org/10.1088/0256-307X/42/7/070713}{Chin. Phys. Lett. \textbf{42}, 070713 (2025)}.


\bibitem{Sun-Nature621} H. L. Sun, M. W. Huo, X. W. Hu, J. Y. Li, Z. J. Liu, Y. F. Han, L. Y. Tang, Z. Q. Mao, P. T. Yang, B. S. Wang, J. G. Cheng, D. X. Yao, G. M. Zhang, and M. Wang, Signatures of superconductivity near 80 K in a nickelate under high pressure,
\href{https://doi.org/10.1038/s41586-023-06408-7}{Nature (London) \textbf{621}, 493 (2023)}.


\bibitem{Hou-CPL40} J. Hou, P. T. Yang, Z. Y. Liu, J. Y. Li, P. F. Shan, L. Ma, G. Wang, N. N. Wang, H. Z. Guo, J. P. Sun, Y. Uwatoko, M. Wang, G. M. Zhang, B. S. Wang, and J. G. Cheng, Emergence of High-Temperature Superconducting Phase in Pressurized La$_3$Ni$_2$O$_7$ Crystals,
\href{https://doi.org/10.1088/0256-307X/40/11/117302}{Chin. Phys. Lett. \textbf{40}, 117302 (2023)}.

\bibitem{Zhang-NatPhy20} Y. N. Zhang, D. J. Su, Y. E. Huang, Z. Y. Shan, H. L. Sun, M. W. Huo, K. X. Ye, J. W. Zhang, Z. H. Yang, Y. K. Xu, Y. Su, R. Li, M. Smidman, M. Wang, L. Jiao, and H. Q. Yuan, High-temperature superconductivity with zero resistance and strange-metal behaviour in La$_3$Ni$_2$O$_{7-\delta}$, \href{https://doi.org/10.1038/s41567-024-02515-y}{Nat. Phys. \textbf{20}, 1269 (2024)}.

\bibitem{Wang-CPL29} Q. Y. Wang, Z. Li, W. H. Zhang, Z. C. Zhang, J. S. Zhang, W. Li, H. Ding, Y. B. Ou, P. Deng, K. Chang, J. Wen, C. L. Song, K. He, J. F. Jia, S. H. Ji, Y. Y. Wang, L. L. Wang, X. Chen, X. C. Ma, and Q. K. Xue, Interface-Induced High-Temperature Superconductivity in Single Unit-Cell FeSe Films on SrTiO$_3$, \href{https://doi.org/10.1088/0256-307X/29/3/037402}{Chin. Phys. Lett. \textbf{29}, 037402 (2012)}.

\bibitem{Liu-PRB85} K. Liu, Z. Y. Lu, and T. Xiang, Atomic and electronic structures of FeSe monolayer and bilayer thin films on SrTiO$_3$ (001): First-principles study, \href{https://doi.org/10.1103/PhysRevB.85.235123}{Phys. Rev. B \textbf{85}, 235123 (2012)}.

\bibitem{Zhang-PRL135} M. D. Zhang, R. C. Shi, R. Wu, X. T. Jiao, M. X. Shi, J. X. Liu, Y. B. Wang, W. F. Dong, C. Ding, T. X. Qin, H. Y. Liu, L. L. Wang, Z. Y. Zhang, P. Gao, Q. K. Xue, and Q. H. Xiong, Direct Evidence of Interfacial Coherent Electron-Phonon Coupling in Single-Unit-Cell FeSe Film on Nb-Doped SrTiO$_3$, \href{https://doi.org/10.1103/qxfc-khzf}{Phys. Rev. Lett. \textbf{135}, 126903 (2025)}.

\bibitem{Ko-Nature638} E. K. Ko, Y. J. Yu, Y. D. Liu, L. Bhatt, J. R. Li, V. Thampy, C. T. Kuo, B. Y. Wang, Y. Lee, K. Lee, J. S. Lee, B. H. Goodge, D. A. Muller, and H. Y. Hwang, Signatures of ambient pressure superconductivity in thin film La$_3$Ni$_2$O$_7$,
\href{https://doi.org/10.1038/s41586-024-08525-3}{Nature (London) \textbf{638}, 935 (2025)}.


\bibitem{Zhou-Nature640} G. D. Zhou, W. Lv, H. Wang, Z. H. Nie, Y. Q. Chen, Y. Y. Li, H. L. Huang, W. Q. Chen, Y. J. Sun, Q. K. Xue, and Z. Y. Chen,
Ambient-pressure superconductivity onset above 40 K in (La,Pr)$_3$Ni$_2$O$_7$ films, \href{https://doi.org/10.1038/s41586-025-08755-z}{Nature (London) \textbf{640}, 641 (2025)}.


\bibitem{Liu-NM24} Y. D. Liu, E. K. Ko, Y. J. Tarn, L. Bhatt, J. R. Li, V. Thampy, B. H. Goodge, D. A. Muller, S. Raghu, Y. J. Yu, and H. Y. Hwang,
Superconductivity and normal-state transport in compressively strained La$_2$PrNi$_2$O$_7$ thin films,
\href{https://doi.org/10.1038/s41563-025-02258-y}{Nat. Mater. \textbf{24}, 1221 (2025)}.

\bibitem{Li-NSR12} P. Li, G. D. Zhou, W. Lv, Y. Y. Li, C. M. Yue, H. L. Huang, L. Z. Xu, J. C. Shen, Y. Miao, W. H. Song, Z. H. Nie, Y. Q. Chen, H. Wang, W. Q. Chen, Y. B. Huang, Z. H. Chen, T. Qian, J. H. Lin, J. F. He, Y. J. Sun, Z. Y. Chen, and Q. K. Xue,
Angle-resolved photoemission spectroscopy of superconducting (La,Pr)$_3$Ni$_2$O$_7$/SrLaAlO$_4$ heterostructures, \href{https://doi.org/10.1093/nsr/nwaf205}{Natl. Sci. Rev. \textbf{12}, nwaf205 (2025)}.


\bibitem{Shi-CPL42} H. L. Shi, Z. H. Huo, G. L. Li, H. Ma, T. Cui, D. X. Yao, and D. F. Duan,
The Effect of Carrier Doping and Thickness on the Electronic Structures of La$_3$Ni$_2$O$_7$ Thin Films,
\href{https://doi.org/10.1088/0256-307X/42/8/080708}{Chin. Phys. Lett. \textbf{42}, 080708 (2025)}.

\bibitem{Deng-QuantumFront4} Z. X. Deng, H. Yan, B. J. Wu, Y. J. Li, J. C. Zhang, Z. W. Zhu, C. L. Song, X. C. Ma, and Q. K. Xue,
Tensile strain effect on superconductivity and thermally activated vortex motion in Sr$_{1-x}$Eu$_x$CuO$_{2+y}$ thin films,
\href{https://doi.org/10.1007/s44214-025-00090-8}{Quantum Front. \textbf{4}, 18 (2025)}.


\bibitem{PereiraJunior-FlatChem37} M. L. Pereira J\'unior, W. F. da Cunha, W. F. Giozza, R. T. de Sousa Junior, and L. A. Ribeiro Junior,
Irida-graphene: A new 2D carbon allotrope,
\href{https://doi.org/10.1016/j.flatc.2023.100469}{FlatChem \textbf{37}, 100469 (2023)}.


\bibitem{Novoselov-Science306} K. S. Novoselov, A. K. Geim, S. V. Morozov, D. Jiang, Y. Zhang, S. V. Dubonos, I. V. Grigorieva, and A. A. Firsov,
Electric Field Effect in Atomically Thin Carbon Films,
\href{https://doi.org/10.1126/science.1102896}{Science \textbf{306}, 666 (2004)}.

\bibitem{Marsden-JOC70} J. A. Marsden and M. M. Haley, Carbon Networks Based on Dehydrobenzoannulenes. 5. Extension of Two-Dimensional Conjugation in Graphdiyne Nanoarchitectures, \href{https://doi.org/10.1021/jo050926v}{J. Org. Chem. \textbf{70}, 10213 (2005)}.

\bibitem{Li-CC46} G. X. Li, Y. L. Li, H. B. Liu, Y. B. Guo, Y. J. Li, and D. B. Zhu, Architecture of graphdiyne nanoscale films,
\href{https://doi.org/10.1039/B922733D}{Chem. Commun. \textbf{46}, 3256 (2010)}.



\bibitem{Li-Carbon136} Q. D. Li, Y. Li, Y. Chen, L. L. Wu, C. F. Yang, and X. L. Cui, Synthesis of $\gamma$-graphyne by mechanochemistry and its electronic structure, \href{https://doi.org/10.1016/j.carbon.2018.04.081}{Carbon \textbf{136}, 248 (2018)}.


\bibitem{Desyatkin-JACS144} V. G. Desyatkin, W. B. Martin, A. E. Aliev, N. E. Chapman, A. F. Fonseca, D. S. Galv\~ao, E. R. Miller, K. H. Stone, Z. Wang, D. Zakhidov, F. T. Limpoco, S. R. Almahdali, S. M. Parker, R. H. Baughman, and V. O. Rodionov, Scalable Synthesis and Characterization of Multilayer $\gamma$-Graphyne, New Carbon Crystals with a Small Direct Band Gap,
\href{https://doi.org/10.1021/jacs.2c06583}{J. Am. Chem. Soc. \textbf{144}, 17999 (2022)}.

\bibitem{Fan-Science372} Q. T. Fan, L. H. Yan, M. W. Tripp, O. Krej\v{c}\'i, S. Dimosthenous, S. R. Kachel, M. Chen, A. S. Foster, U. Koert, P. Liljeroth, and J. M. Gottfried, Biphenylene network: A nonbenzenoid carbon allotrope,
\href{https://doi.org/10.1126/science.abg4509}{Science \textbf{372}, 852 (2021)}.

\bibitem{Liu-PRB76} F. Liu, P. B. Ming, and J. Li, Ab initio calculation of ideal strength and phonon instability of graphene under tension,
\href{https://doi.org/10.1103/PhysRevB.76.064120}{Phys. Rev. B \textbf{76}, 064120 (2007)}.


\bibitem{Penev-NL16} E. S. Penev, A. Kutana, and B. I. Yakobson, Can Two-Dimensional Boron Superconduct?,
\href{https://doi.org/10.1021/acs.nanolett.6b00070}{Nano Lett. \textbf{16}, 2522 (2016)}.

\bibitem{Gao-PRB95} M. Gao, Q. Z. Li, X. W. Yan, and J. Wang, Prediction of phonon-mediated superconductivity in borophene,
\href{https://doi.org/10.1103/PhysRevB.95.024505}{Phys. Rev. B \textbf{95}, 024505 (2017)}.

\bibitem{Gao-PRB100} M. Gao, X. W. Yan, J. Wang, Z. Y. Lu, and T. Xiang, Electron-phonon coupling in a honeycomb borophene grown on Al(111) surface,
\href{https://doi.org/10.1103/PhysRevB.100.024503}{Phys. Rev. B \textbf{100}, 024503 (2019)}.

\bibitem{Kamal-PRB91} C. Kamal and M. Ezawa, Arsenene: Two-dimensional buckled and puckered honeycomb arsenic systems,
\href{https://doi.org/10.1103/PhysRevB.91.085423}{Phys. Rev. B \textbf{91}, 085423 (2015)}.

\bibitem{Kong-CPB27} X. Kong, M. Gao, X. W. Yan, Z. Y. Lu, and T. Xiang, Superconductivity in electron-doped arsenene,
\href{https://doi.org/10.1088/1674-1056/27/4/046301}{Chin. Phys. B \textbf{27}, 046301 (2018)}.


\bibitem{Sahin-PRB80} H. {\c{S}}ahin, S. Cahangirov, M. Topsakal, E. Bekaroglu, E. Akturk, R. T. Senger, and S. Ciraci, Monolayer honeycomb structures of group-IV elements and III-V binary compounds: First-principles calculations, \href{https://doi.org/10.1103/PhysRevB.80.155453}{Phys. Rev. B \textbf{80}, 155453 (2009)}.

\bibitem{Maultzsch-PRL92} J. Maultzsch, S. Reich, C. Thomsen, H. Requardt, and P. Ordej\'on, Phonon Dispersion in Graphite, \href{https://doi.org/10.1103/PhysRevLett.92.075501}{Phys. Rev. Lett. \textbf{92}, 075501 (2004)}.

\bibitem{Bonini-PRL99} N. Bonini, M. Lazzeri, N. Marzari, and F. Mauri, Phonon Anharmonicities in Graphite and Graphene, \href{https://doi.org/10.1103/PhysRevLett.99.176802}{Phys. Rev. Lett. \textbf{99}, 176802 (2007)}.

\bibitem{Bekaert-PRB96} J. Bekaert, A. Aperis, B. Partoens, P. M. Oppeneer, and M. V. Milo{\v{s}}evi{\'c}, Evolution of multigap superconductivity in the atomically thin limit: Strain-enhanced three-gap superconductivity in monolayer MgB$_2$, \href{https://doi.org/10.1103/PhysRevB.96.094510}{Phys. Rev. B \textbf{96}, 094510 (2017)}.

\bibitem{Bekaert-PRL123} J. Bekaert, M. Petrov, A. Aperis, P. M. Oppeneer, and M. V. Milo{\v{s}}evi{\'c}, Hydrogen-Induced High-Temperature Superconductivity in Two-Dimensional Materials: The Example of Hydrogenated Monolayer MgB$_2$, \href{https://doi.org/10.1103/PhysRevLett.123.077001}{Phys. Rev. Lett. \textbf{123}, 077001 (2019)}.


\bibitem{Gao-PRB101} M. Gao, X. W. Yan, Z. Y. Lu, and T. Xiang, Strong-coupling superconductivity in LiB$_2$C$_2$ trilayer films, \href{https://doi.org/10.1103/PhysRevB.101.094501}{Phys. Rev. B \textbf{101}, 094501 (2020)}.

\bibitem{Liu-NL23} L. Liu, X. Liu, P. Song, L. Zhang, X. Huang, W. Zhang, Z. Zhang, and Y. Jia, Surface Superconductivity with High Transition Temperatures in Layered Ca$_n$B$_{n+1}$C$_{n+1}$ Films, \href{https://doi.org/10.1021/acs.nanolett.2c05038}{Nano Lett. \textbf{23}, 1924 (2023)}.


\bibitem{Zhang-PRB110} Y. M. Zhang, J. Y. Chen, J. Hao, M. L. Xu, and Y. W. Li, Conventional high-temperature superconductivity in $\sigma$-band driven metallized two-dimensional metal borocarbides, \href{https://doi.org/10.1103/PhysRevB.110.064513}{Phys. Rev. B \textbf{110}, 064513 (2024)}.

\bibitem{Chen-CPL41} W. X. Chen, Z. T. Liu, Z. H. Huo, G. Y. Dong, J. L. Cai, and D. F. Duan, High-Temperature Phonon-Mediated Superconductivity with $T_c$ above 100 K in Monolayer Na(BC)$_2$ and K(BC)$_2$, \href{https://doi.org/10.1088/0256-307X/41/11/117403}{Chin. Phys. Lett. \textbf{41}, 117403 (2024)}.


\bibitem{Jin-PRB101} X. T. Jin, X. W. Yan, and M. Gao, First-principles calculations of monolayer hexagonal boron nitride: Possibility of superconductivity, \href{https://doi.org/10.1103/PhysRevB.101.134518}{Phys. Rev. B \textbf{101}, 134518 (2020)}.


\bibitem{Liu-PRR6} H. D. Liu, B. T. Wang, Z. G. Fu, H. Y. Lu, and P. Zhang, Three-gap superconductivity with $T_c$ above 80 K in hydrogenated 2D monolayer LiBC, \href{https://doi.org/10.1103/PhysRevResearch.6.033241}{Phys. Rev. Res. \textbf{6}, 033241 (2024)}.


\bibitem{Ku-PRB107} R. Q. Ku, L. Yan, J. G. Si, S. Y. Zhu, B. T. Wang, Y. D. Wei, K. J. Pang, W. Q. Li, and L. J. Zhou, Ab initio investigation of charge density wave and superconductivity in two-dimensional Janus $2H/1T$-MoSH monolayers, \href{https://doi.org/10.1103/PhysRevB.107.064508}{Phys. Rev. B \textbf{107}, 064508 (2023)}.

\bibitem{Cao-Nature556}  Y. Cao, V. Fatemi, A. Demir, S. Fang, S. L. Tomarken, J. Y. Luo, J. D. Sanchez-Yamagishi, K. Watanabe, T. Taniguchi, E. Kaxiras, R. C. Ashoori, and P. Jarillo-Herrero, Correlated insulator behaviour at half-filling in magic-angle graphene superlattices, \href{https://doi.org/10.1038/nature26154}{Nature \textbf{556}, 80 (2018)}.

\bibitem{Si-PRL111} C. Si, Z. Liu, W. H. Duan, and F. Liu, First-Principles Calculations on the Effect of Doping and Biaxial Tensile Strain on Electron-Phonon Coupling in Graphene, \href{https://doi.org/10.1103/PhysRevLett.111.196802}{Phys. Rev. Lett. \textbf{111}, 196802 (2013)}.

\bibitem{Gu-CPL36} Q. Y. Gu, D. Y. Xing, and J. Sun, Superconducting Single-Layer T-Graphene and Novel Synthesis Routes, \href{https://doi.org/10.1088/0256-307X/36/9/097401}{Chin. Phys. Lett. \textbf{36}, 097401 (2019)}.

\bibitem{Kang-PRB99} Y. T. Kang, C. Lu, F. Yang, and D. X. Yao, Single-orbital realization of high-temperature $s^\pm$ superconductivity in the square-octagon lattice, \href{https://doi.org/10.1103/PhysRevB.99.184506}{Phys. Rev. B \textbf{99}, 184506 (2019)}.

\bibitem{Yao-CPL42} Y. Yao, J. Li, J. Ye, F. Yang, and D. X. Yao, Electric Field Induced Superconductivity in Bilayer Octagraphene, \href{https://doi.org/10.1088/0256-307X/42/6/067504}{Chin. Phys. Lett. \textbf{42}, 067504 (2025)}.

\bibitem{Ye-CPL40} J. C. Ye, J. Li, D. Y. Zhong, and D. X. Yao, Possible Superconductivity in Biphenylene, \href{https://doi.org/10.1088/0256-307X/40/7/077401}{Chin. Phys. Lett. \textbf{40}, 077401 (2023)}.

\bibitem{Wan-EurophysLett104} W. H. Wan, Y. F. Ge, F. Yang, and Y. G. Yao, Phonon-mediated superconductivity in silicene predicted by first-principles density functional calculations, \href{https://doi.org/10.1209/0295-5075/104/36001}{Europhys. Lett. \textbf{104}, 36001 (2013)}.


\bibitem{Ge-NewJPhys17} Y. F. Ge, W. H. Wan, F. Yang, and Y. G. Yao, The strain effect on superconductivity in phosphorene: a first-principles prediction, \href{https://doi.org/10.1088/1367-2630/17/3/035008}{New J. Phys. \textbf{17}, 035008 (2015)}.


\bibitem{Zhang_CPB32} L. P. Zhang, Z. Y. Xu, X. J. Li, X. Zhang, M. Y. Qin, R. Z. Zhang, J. Xu, W. X. Cheng, J. Yuan, H. B. Wang, A. V. Silhanek, B. Y. Zhu, J. Miao, and K. Jin, Cascade excitation of vortex motion and reentrant superconductivity in flexible Nb thin films, \href{https://doi.org/10.1088/1674-1056/acac16}{Chin. Phys. B \textbf{32}, 047302 (2023)}.


\bibitem{Guo-Science306} Y. Guo, Y. F. Zhang, X. Y. Bao, T. Z. Han, Z. Tang, L. X. Zhang, W. G. Zhu, E. G. Wang, Q. Niu, Z. Q. Qiu, J. F. Jia, Z. X. Zhao, and Q. K. Xue, Superconductivity Modulated by Quantum Size Effects, \href{https://doi.org/10.1126/science.1105130}{Science \textbf{306}, 1915 (2004)}.

\bibitem{Nishio-PRL101}  T. Nishio, T. An, A. Nomura, K. Miyachi, T. Eguchi, H. Sakata, S. Lin, N. Hayashi, N. Nakai, M. Machida, and Y. Hasegawa, Superconducting Pb Island Nanostructures Studied by Scanning Tunneling Microscopy and Spectroscopy, \href{https://doi.org/10.1103/PhysRevLett.101.167001}{Phys. Rev. Lett. \textbf{101}, 167001 (2008)}.


\bibitem{Zhang-PRL114} H. M. Zhang, Y. Sun, W. Li, J. P. Peng, C. L. Song, Y. Xing, Q. H. Zhang, J. Q. Guan, Z. Li, Y. F. Zhao, S. H. Ji, L. L. Wang, K. He, X. Chen, L. Gu, L. S. Ling, M. L. Tian, L. Li, X. C. Xie, J. P. Liu, H. Yang, Q. K. Xue, J. Wang, and X. Ma, Detection of a Superconducting Phase in a Two-Atom Layer of Hexagonal Ga Film Grown on Semiconducting GaN(0001), \href{https://doi.org/10.1103/PhysRevLett.114.107003}{Phys. Rev. Lett. \textbf{114}, 107003 (2015)}.


\bibitem{Giannozzi-JPhysCondensMatter21} P. Giannozzi, S. Baroni, N. Bonini, M. Calandra, R. Car, C. Cavazzoni, D. Ceresoli, G. L. Chiarotti, M. Cococcioni, I. Dabo, A. Dal Corso, S. de Gironcoli, S. Fabris, G. Fratesi, R. Gebauer, U. Gerstmann, C. Gougoussis, A. Kokalj, M. Lazzeri, L. Martin-Samos, N. Marzari, F. Mauri, R. Mazzarello, S. Paolini, A. Pasquarello, L. Paulatto, C. Sbraccia, S. Scandolo, G. Sclauzero, A. P. Seitsonen, A. Smogunov, P. Umari, and R. M. Wentzcovitch, QUANTUM ESPRESSO: a modular and open-source software project for quantum simulations of materials, \href{https://doi.org/10.1088/0953-8984/21/39/395502}{J. Phys.: Condens. Matter \textbf{21}, 395502 (2009)}.


\bibitem{Perdew-PRL77} J. P. Perdew, K. Burke, and M. Ernzerhof, Generalized Gradient Approximation Made Simple, \href{https://doi.org/10.1103/PhysRevLett.77.3865}{Phys. Rev. Lett. \textbf{77}, 3865 (1996)}.

\bibitem{Hamann-PRB88} D. R. Hamann, Optimized norm-conserving Vanderbilt pseudopotentials, \href{https://doi.org/10.1103/PhysRevB.88.085117}{Phys. Rev. B \textbf{88}, 085117 (2013)}.


\bibitem{Methfessel-PRB40} M. Methfessel and A. T. Paxton, High-precision sampling for Brillouin-zone integration in metals, \href{https://doi.org/10.1103/PhysRevB.40.3616}{Phys. Rev. B \textbf{40}, 3616 (1989)}.

\bibitem{Baroni-RMP73} S. Baroni, S. de Gironcoli, A. Dal Corso, and P. Giannozzi, Phonons and related crystal properties from density-functional perturbation theory, \href{https://doi.org/10.1103/RevModPhys.73.515}{Rev. Mod. Phys. \textbf{73}, 515 (2001)}.


\bibitem{Pizzi-JPCM32} G. Pizzi, V. Vitale, R. Arita, S. Bl\"ugel, F. Freimuth, G. G\'eranton, M. Gibertini, D. Gresch, C. Johnson, T. Koretsune, J. Iba\~{n}ez-Azpiroz, H. Lee, J. M. Lihm, D. Marchand, A. Marrazzo, Y. Mokrousov, J. I. Mustafa, Y. Nohara, Y. Nomura, L. Paulatto, S. Ponc\'e, T. Ponweiser, J. Qiao, F. Th\"ole, S. S. Tsirkin, M. Wierzbowska, N. Marzari, D. Vanderbilt, I. Souza, A. A. Mostofi, and J. R. Yates, Wannier90 as a community code: new features and applications, \href{https://doi.org/10.1088/1361-648X/ab51ff}{J. Phys.: Condens. Matter \textbf{32}, 165902 (2020)}.


\bibitem{Lee-npjCM9} H. Lee, S. Ponc\'e, K. Bushick, S. Hajinazar, J. Lafuente-Bartolome, J. Leveillee, C. Lian, J. M. Lihm, F. Macheda, H. Mori, H. Paudyal, W. H. Sio, S. Tiwari, M. Zacharias, X. Zhang, N. Bonini, E. Kioupakis, E. R. Margine, and F. Giustino, Electron-phonon physics from first principles using the EPW code, \href{https://doi.org/10.1038/s41524-023-01107-3}{npj Comput. Mater. \textbf{9}, 156 (2023)}.

\end{references}
\end{document}